\documentclass[12pt,a4paper]{article}
\usepackage[utf8]{inputenc}
\usepackage{amsmath}
\usepackage{amsfonts}
\usepackage{amssymb}
\usepackage{graphicx}
\usepackage{tikz-cd}
\usepackage{hyperref}

\graphicspath{ {images/} }

\usepackage{bbm}
\usepackage{bm}
\usepackage[left=2cm,right=2cm,top=2cm,bottom=2cm]{geometry}
\usepackage{mathtools}
\usepackage[round]{natbib}
\makeatletter 
\renewcommand\@biblabel[1]{#1.} 
\makeatother %

\newtheorem{theorem}{Theorem}
\newtheorem{lemma}{Lemma}

\DeclarePairedDelimiter{\ceil}{\lceil}{\rceil}
\DeclarePairedDelimiter\floor{\lfloor}{\rfloor}
\title{Model-aware Quantile Regression for Discrete Data}
\usepackage{authblk}

\author[1]{Tullia Padellini}
\author[2]{ H\aa vard Rue}
\affil[1]{\small{Dipartimento di Scienze Statistiche -- Sapienza Università di Roma}}
\affil[2]{\small{CEMSE Division -- King Abdullah University of Science and Technology}}

\makeindex

\begin{document}
\maketitle
\abstract{
 Quantile regression relates the quantile of the response to a
    linear predictor. For a discrete response distributions, like the
    Poission, Binomial and the negative Binomial, this approach is not
    feasible as the quantile function is not bijective. We argue to
    use a continuous model-aware interpolation of the quantile
    function, allowing for proper quantile inference while retaining
    model interpretation. This approach allows for proper uncertainty
    quantification and mitigates the issue of quantile crossing. Our
    reanalysis of hospitalisation data considered in
    \cite{Congdon2017} shows the advantages of our proposal as well as
    introducing a novel method to exploit quantile regression in the
    context of disease mapping.
 }
\section{Introduction}

Quantile regression is a supervised technique aimed at modelling the
quantiles of the conditional distribution of some response variable,
introduced by \cite{Koenker1978}. Mean regression is concerned with
modelling conditional expectations, whereas quantile regression is
especially useful when the tails of the distribution are of interest,
as for example when the focus is on extreme behaviour, or when
covariates may not affect the whole population uniformly.

Let $Q_{\alpha}(Y_i\mid X_i)$ be the quantile of level $\alpha$ of the
conditional distribution of the $i$th observation $Y_i\mid X_i$, in a
sample of size $n$. Point estimates for quantile regression models of
the form
\begin{equation}\label{eq:quant_model}
    Q_{\alpha}(Y_i\mid X_i) = X_i^t \beta_{\alpha}  \quad (i = 1,\dots , n)
\end{equation}
are typically obtained by minimising the following empirical risk:
\begin{equation}
    \label{eq:beta_check}
    \widehat{\beta}_{\alpha} =
    \arg\min_{\beta_{\alpha}} \sum_{i = 1} ^n \rho_{\alpha}
    (Y_i - X_i ^ t \beta_{\alpha}. )
\end{equation}
Here $\rho_\alpha(x) = x(\alpha-\mathbb{I}{\{x<0\}})$, with
$\mathbb{I}$ being the indicator function, is known as the
\emph{check} loss function. As the optimisation problem
in~\eqref{eq:beta_check} does not depend on the distribution of the
response variable, quantile regression is typically considered a
model-free technique. Even likelihood based methods, including
Bayesian procedures, for which a model assumption on the dependent
variables is needed, do not often exploit the generating distribution
of the response variable, but make use of working likelihoods instead.
The classical model assumption in quantile regression is that of the
response being generated by an Asymmetric Laplace Distribution
\citep{Yu2001,Yue2011}, which, rather than representing an hypothesis
on the generating mechanism of the data, is motivated by the fact the
resulting Maximum Likelihood estimator coincides with the optimum
of~\eqref{eq:beta_check}.

These approaches fail when the response variable is discrete. The
Asymmetric Laplace distribution assumption, in fact, implies that the
observations are continuous, while in the approach based on the direct
optimisation of~ \eqref{eq:beta_check}, inferential procedure beyond
point estimations are compromised by the non-differentiability of the
objective function, which, together with the points of positive mass
of one of the variables involved in the optimisation problem, make it
challenging to derive an asymptotic distribution for the sample
quantiles \citep{Machado2005}. Quantile inference for discrete data
cannot thus be carried out directly, but it is necessary to enforce
additional smoothing on the response variable, and model a continuous
approximation of it instead.

In most of the quantile regression literature, such approximation is
obtained by means of \emph{jittering}, i.e. perturbing the discrete
distribution by adding continuous and bounded noise, as first proposed
by \cite{Machado2005}. When the noise taken to be an Uniform random variable between $0$ and $1$, jittering can be thought of as an interpolation
strategy, as the following one-to- one relationship between the
quantiles of the jittered response $Y_i\mid X_i$ and those of the
original variable of interest $Y^*_i\mid X_i$ holds for each
observation in the sample:
\[Q_{\alpha}(Y^*_i\mid X_i) = \ceil{ Q_{\alpha}(Y_i\mid X_i) -1} \quad
    (i = 1, \dots, n).\]
Regression models can be specified and fitted for
$Q_{\alpha}(Y_i\mid X_i)$, since it is a continuous function, however,
the dependence of the distribution of $Y_i\mid X_i$ on the
distribution of the arbitrarily chosen noise variable, may hinder the
interpretation of $Q_{\alpha}(Y_i\mid X_i)$ as a continuous version of
$Q_{\alpha}(Y^*_i\mid X_i)$.
 
We argue to use a model-aware interpolation strategy for building
continuous quantile function to be exploited in quantile regression.
Our continuous interpolation is tailored to the true distribution of
the discrete response and retains the model interpretation of the
discrete counterpart, thus providing a stronger justification to the
modelling of the continuous quantiles as a proxy of the discrete ones.
Additionally, our interpolation scheme also overcomes two
more drawbacks of jittering, namely the fact that the new continuous
variable is smooth over the entire support and it does not depend on
specific realizations of the noise.

In order to fully exploit the distributional assumptions required for
the approximation of discrete response variables, we propose a new
approach to quantile regression, based on the direct modelling of the
quantiles of the generating distribution; in doing so, we address the
lack of generating likelihoods in quantile regression. This proposal
allows us to extend the Generalized Linear (Mixed) Model framework to
quantile regression by redefining the link function. This formulation
not only recast quantile regression in a more cohesive setting
and overcomes the fragmentation of quantile regression literature, but
it is also key to an efficient and ready to use fitting procedure, as
the connection allows to estimate the model using \texttt{R-INLA}
\citep{Rue2009,art527,Rue2017,art643,book126}. We show that by
exploiting the response distribution in the fitting procedures, it is
possible to mitigate the phenomenon of quantile crossing, to provide a
unified framework for quantile regression and to obtain proper
uncertainty quantification in Bayesian settings. 

\section{Model-Based Quantile
    Regression}\label{sec:QuantileRegression}

As opposed to mean regression, where generalization of the basic
linear model heavily rely on the response distribution, in quantile
regression, with the noticeable exception of \cite{Noufaily2013,art642,art650},
generating models are rarely considered. The likelihood assumption
commonly found in quantile regression, the Asymmetric Laplace
Distribution, is not motivated by the shape of the data. The use of
the Asymmetric Laplace Distribution with respect to a proper model may
seem appealing due to the apparently weak modeling assumption on the
response. However, adopting the Asymmetric Laplace Distribution
imposes several restriction that may not be obvious nor desirable in
applications: the skewness of the density is fully determined when a
specific percentile is chosen, the density is symmetric when
\(\alpha=0.5\) and the mode of the error distribution is at \(0\)
regardless of \(\alpha\), which results in rigid error density tails
for extreme percentiles \citep{Yan2017}.

The limitations of using a working likelihood are even more critical
in the Bayesian framework, where the lack of a generating likelihood
implies that the validity of posterior inference is no longer
guaranteed by Bayes Theorem. As shown by \cite{Yang2016}, the scale
parameter of the Asymmetric Laplace Distribution affects the posterior
variance, despite not having any impact on the quantile itself, and
although they provide a corrected adjusted variance, their result is
only asymptotically valid.

We reject altogether the use of a working likelihood in favor of the
true generating model and we argue to use a \emph{model-based quantile
    regression}, which exploits the shape of the conditional
distribution to link the covariates of interest to the distribution
parameter. This approach is general enough to be adopted in any
inferential paradigm, however it is especially appealing in the
Bayesian framework since allows for proper uncertainty quantification. Our
fitting procedure can be formalized in two steps.

\begin{description}
\item[Step1 - Modelling.]
    Assume that for each unit $i$, $Y_i\mid X_i$ has a known
    continuous distribution $F(y_i;\theta_i)$, where $\theta_i$ is the
    model parameter, for simplicity $\theta_i \in \mathbb{R}$. For each
    $\alpha$, the quantile of $Y_i\mid X_i$,
    \(Q_{\alpha}(Y_i\mid X_i)\) is modelled as
    \begin{equation}
        Q_{\alpha}(Y_i\mid X_i) = g(\eta^{\alpha}_i) \quad (i = 1, \dots, n)
    \end{equation}
    where $g$ is a well behaved link function and $\eta_i^{\alpha}$ is
    the linear predictor, which depends on the level $\alpha$ of the
    quantile. No restriction needs to be placed on the linear
    predictor, which can include fixed as well as random effect. Our
    approach is thus flexible enough to include parametric or semi
    parametric models, where the interest may lay in assessing the
    difference in the impact that the covariate may have at different
    levels of the distribution, as well as fully non parametric
    models, where the focus is on prediction.
\end{description}
\begin{description}
\item[Step 2 - Mapping.]
    The quantile $Q_{\alpha}(Y_i\mid X_i)$ is mapped to the parameter
    $\theta_i$ as
    \begin{equation} \theta_i = h(Q_{\alpha}(Y_i\mid X_i), \alpha)
        \quad (i = 1, \dots, n)
    \end{equation}
    where the function $h$ must be invertible to ensure the
    identifiability of the model and explicitly depends on the
    quantile level $\alpha$. The map $h$ gives us a first
    interpretation of model-based quantile regression as a
    reparametrization of the generating likelihood function
    $F(y_i; \theta_i)$, in terms of its quantiles, i.e.
    $F(y_i; Q_{\alpha}(Y_i\mid X_i) = h^{-1}(\theta_i, \alpha) )$.
\end{description}
By linking the quantiles of the generating distribution to its
parameter $\theta_i$, we are indirectly modeling $\theta_i$ as well,
hence we are implicitly building a connection between quantile
regression and Generalized Linear (Mixed) Models. The modeling and
mapping steps can be considered as a way to define the link function,
in the Generalized Linear Models sense, as the composition
\(\theta_i = h(g(\eta_i))\). This allows us to rephrase quantile
regression as a new link function in a standard Generalized Linear
Models problem. Drawing a path from Generalized Linear Models to
quantile regression is instrumental in the fitting however the pairing
is only formal: coefficients and random effect retain different
interpretations. From a computational standpoint, the main advantage
of coupling quantile regression to Generalized Linear Models is that
this new class of models can be fitted in standard software packages like \texttt{R-INLA}, which
allows for both flexibility in the model definition and efficiency in
their fitting. Extensions to the case of multivariate parameters, i.e.
\(\theta_i \in \mathbb{R}^d \), with \(d>1\), are also possible, and
require all the components of the model parameter \(\theta_i\) to be
redefined as a function of the quantiles.

\section{Poisson Data}\label{sec:discrete_data}

Quantile regression cannot be defined for discrete responses,
hence the standard practice is to model a continuous approximation of
the quantile function. When such approximation is obtained via mean of
jittering as in \cite{Machado2005}, the dependence of the shape of the
continuous quantiles on the distribution of the noise variable may
hinder the interpretation of the jittered quantile.

\begin{figure}
    \centering \includegraphics[width=.9\textwidth]{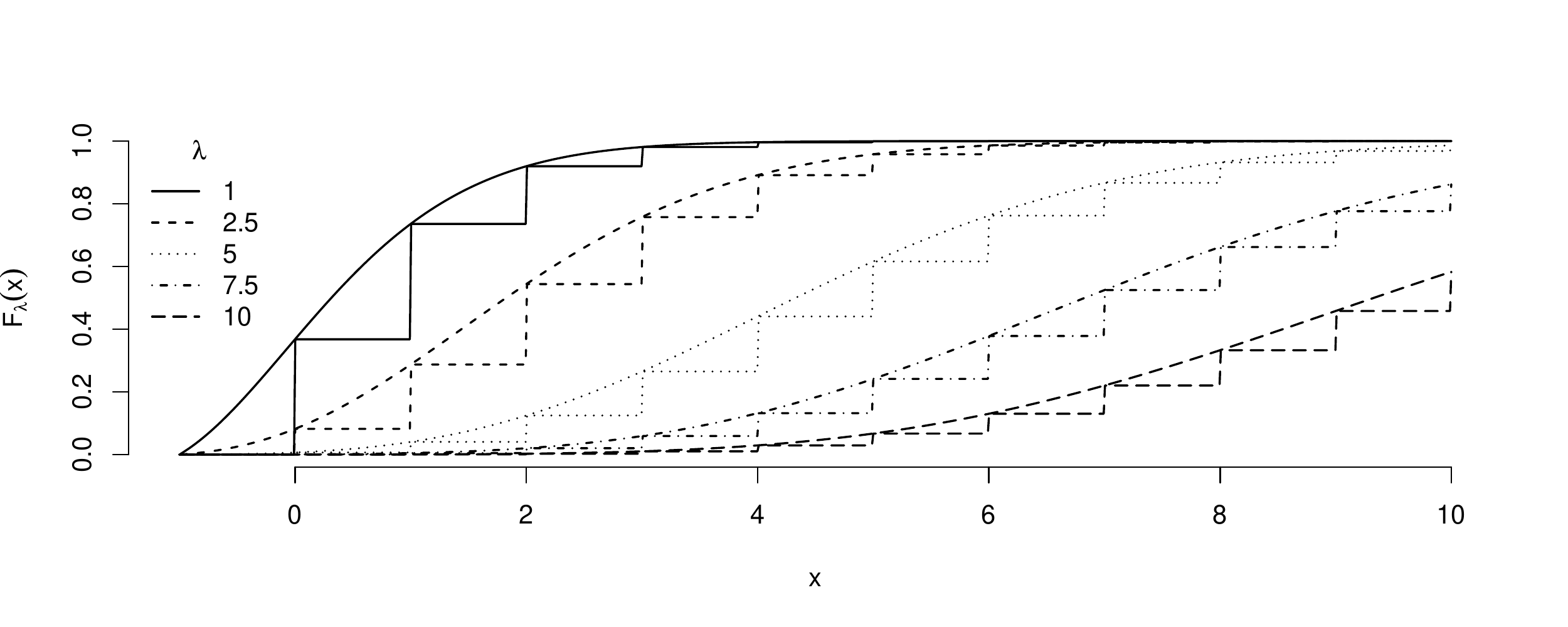}
    \includegraphics[width=.9\textwidth]{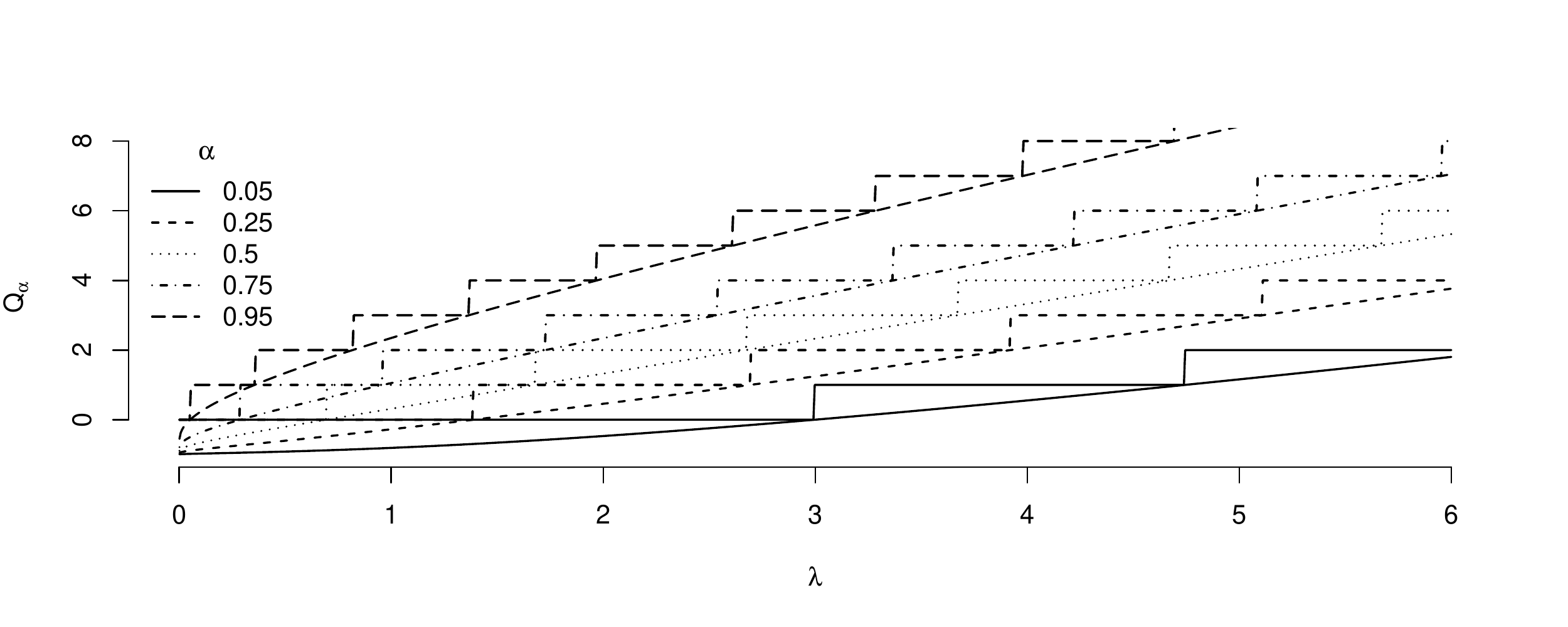}
    \caption{Top: c.d.f. of the discrete and continuous Poisson
        distribution for several values of $\lambda$. Bottom: quantile
        function of the discrete and continuous Poisson distribution
        as a function of $\lambda$.}
    \label{fig:poisson}
\end{figure}

We build the continuous approximation by providing a model-aware
strategy, resulting in a stronger justification of the new quantile
functions as a continuous version of the discrete ones. Inspired by
\cite{Ilienko2013}, we focus on random variables satisfying the
following assumption, for which the extension to the continuous case
is immediate.\\

\textbf{Assumption} The cumulative distribution of the
    discrete random variable $X^*$ admits the following representation
    \begin{equation}\label{class1}
        F_{X^*}(x; \theta) = k(\floor{x}, \theta)
    \end{equation} 
    where $k$ is a continuous function in the first argument.\\
    \\*
\noindent With this assumption, the continuous interpolation can be defined by
removing the floor operator, and the function $k(x, \theta)$ is the
cumulative distribution function of $X$, a continuous version of
$X^*$. Note that, for all integer $x$
\begin{equation} \label{class2} k(\floor{x}, \theta)= k(x, \theta).
\end{equation} 
Incidentally, this way of making variables continuous is well embedded
in statistical culture: even the discrete and continuous version of
the Uniform distribution are related in a similar manner.

\subsection{Continuous Poisson}

The class of distribution defined by~\eqref{class1} and
\eqref{class2} is broad enough to include the three discrete
distribution most frequently encountered in applications: Poisson,
Binomial and Negative Binomial. We explore in detail the Poisson case;
the other two are trivial extensions.

We write the cumulative density function for the Poisson as the ratio
of Incomplete and Regular Gamma function:
\begin{equation}
    X^*\sim \text{Poisson}(\lambda) \quad \quad  F_{X^*}(x) =
    \frac{\Gamma( \floor{x}+1, \lambda)}{\Gamma(\floor{x}+1)}\quad x\geq 0
\end{equation}
where
\[\Gamma(x, \lambda) = \int _{\lambda} ^ \infty e^ {-s} s^{x-1}
    \text{d}s\] is the upper incomplete Gamma function. As in
\cite{Ilienko2013}, extending the Poisson distribution to the
continuous case from this formulation is now immediate:
\begin{equation}
    X \sim \text{Continuous Poisson}(\lambda) \quad \quad F_{X}(x) =
    \frac{\Gamma( x+1, \lambda)}{\Gamma(x+1)} \quad x> -1.
    \label{eq:cont_poisson}
\end{equation}
where the domain has been extended from $x\geq 0$ to $x>-1$. Following
\cite{Ilienko2013}, it follows directly that $F_{X}(x)$ is a well
defined distribution function, in the sense that it is non-decreasing
in $x$, is right-continuous, $F_{X}(-\infty) = 0$ and
$F_{X}(\infty) = 1$. The definition in~\eqref{eq:cont_poisson} is
similar to that of \cite{Ilienko2013}, but it is shifted by $1$.
\begin{lemma}\label{lemma:interpolation}
    Let $X^*, X$ be respectively a discrete and continuous Poisson
    random variable, both with parameter $\lambda$, then
    $X^* = \ceil{X}$.
\end{lemma}
From an interpretation point of view, the pairing is strengthened by
the fact that the classical result of an Erlang random variable being
the waiting time between occurrences of a Poisson homogeneous process,
can be extended to the Continuous Poisson case. This can be trivially
shown by replacing the Erlang distribution with the Gamma
distribution, which are the same distribution with respectively
discrete and continuous parameters.

\subsection{Quantile Regression for Poisson Data}

Quantile regression for Poisson data can be performed by assuming
that, conditionally on covariates $X_i$, each response $Y^*_i$ is
generated from a Poisson, i.e.
\begin{equation}
    Y^*_i\mid X_i \sim \text{Poisson}(\lambda_i) \quad (i = 1, \dots, n).
\end{equation}
The modelling and mapping step can then be specified as follows:
\begin{description}
\item \textit{Step 1.} Modelling:
    \(Q_{\alpha}(Y_i\mid X_i) = \exp\{\eta_i ^{\alpha}\} \quad (i = 1,
    \dots, n) \)
\item \textit{Step 2.} Mapping:
    \(\lambda_i = \frac{\Gamma^{-1}(Q_{\alpha}(Y_i\mid X_i)
        ^{\alpha}+1, 1-\alpha)}{
        \Gamma(Q_{\alpha}(Y_i\mid X_i) ^{\alpha}+1)} \quad (i = 1,
    \dots, n)\)
\end{description}
where $Y _i\mid X_i\sim \text{Continuous Poisson} (\lambda_i)$. While
modelling quantiles of a continuous approximation implies that the
fitted quantiles curves are not discrete, as a consequence of
Lemma~\ref{lemma:interpolation}, the equivariance property of quantile
guarantees that
\begin{equation}
    Q_{\alpha}(Y_i\mid X_i) =Q_{\alpha}(\ceil{Y^*_i}\mid X_i) =
    \ceil{Q_{\alpha}(Y^*_i\mid X_i)} \quad (i = 1, \dots, n).    
\end{equation}
\subsection{Good Properties - Crossing}

One of the most prominent issues in quantile regression literature is
that estimated quantile curves may intersect when more than one
quantile level is considered. This phenomenon, usually referred to as
\emph{quantile crossing}, which impedes the interpretability of the
results, is a consequence of fitting one quantile curve at the time,
and may be overcame by jointly estimate multiple quantiles, see for
example \cite{Bondell2010a}. Our proposal can also be extended to the
multiple quantile case performing simultaneous estimation by smoothing
over quantile levels through a spline model, as in \cite{Wei2012}.

However, even in the single quantile case, our method is less affected
by this issue by definition, as the presence of a known generating
likelihood informs the fitting mechanism on the other quantiles, hence
reducing the impact of fitting separate models for each quantile.

Moreover, while in general it is not possible to completely avert the crossing, 
as it would mean that quantile curves are
parallel, intersection can be avoided when the domain of the
covariates is bounded.

\begin{figure}
    \includegraphics[width=1\textwidth]{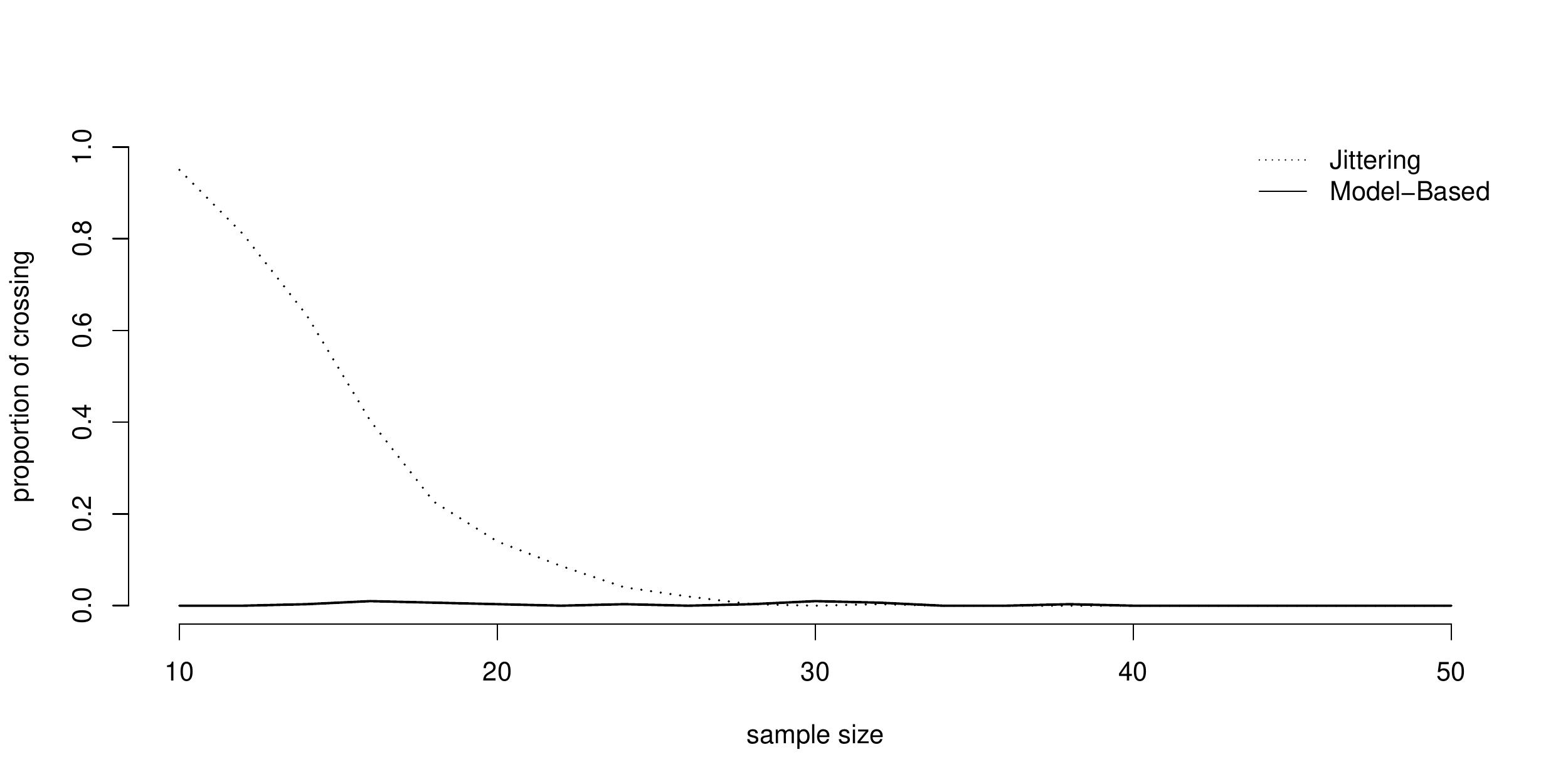} \centering
    \caption{Frequency of at least one quantile crossing for
        model-based (solid line) and jittered (dashed line) quantile
        regression.}
    \label{fig:crossing}
\end{figure}

\begin{lemma}\label{lemma:crossings}
    Let $Y_i ^* \mid X_i$ and $Y_i\mid X_i$ be distributed
    respectively as a discrete and continuous Poisson with parameter
    $\lambda_i$, and let $X_i\in \mathbb{R}^ +$. Consider the model
    $Q_{\alpha}(Y_i\mid X_i) = \exp\{\beta_{\alpha}X_i\}$, then the
    Maximum Likelihood estimator $\hat{\beta_{\alpha}}$ is a non
    decreasing function of $\alpha$.
\end{lemma}
The proof is immediate from the equivariance property of Maximum
Likelihood estimator and the monotonicity of the exponential function.
In the jittering framework, even an apparently restrictive setting as
the above can be troublesome. Figure~\ref{fig:crossing} shows the
frequency of at least one crossing over $300$ datasets generated from
a Poisson random variable $ \lambda_i = \exp(X_i)$, where the
covariates $X_i$ are simulated from the absolute value of a Normal
random variable with mean $0$ and standard deviation $1.5$. In this
toy example, quantile curves are estimated on an equally spaced grid
of $\alpha = (0.05, \dots, 0.95)$, using both the jittering approach
and our approach. While Lemma~\ref{lemma:crossings} holds only
asymptotically for a Bayesian estimator of $\beta_{\alpha}$, unless we
assume a uniform prior distribution for the coefficient
$\beta_{\alpha}$. Figure~ \ref{fig:crossing} seems to suggest that the
behavior with respect to quantile crossings is good even for small
sample sizes.

\section{Continuous Count distributions}

The Binomial and the Negative Binomial distribution can be similarly
extended to the continuous case. Their cumulative distribution
functions can be expressed as:
\begin{align}
  Y^* \sim \text{Binomial}(n, p) & & &F_{Y^*}(y) = I_{1-p}(n-\floor{y}, \floor{y}+1)\\
  Z^* \sim \text{Negative Binomial}(r, p) & & &F_{Z^*}(z) = I_{1-p}(r, \floor{z}+1)
\end{align}
where $I_{x}(a, b)$ is the \emph{regularized incomplete Beta function}
defined as:
\begin{align}
  I_{x}(a, b) = \frac{\text{B}(a, b, x)}{\text{B}(a, b)}
  &&\text{with} \quad& \text{B}(a, b, x) = \int_x ^1 t^a(1-t)^{b-1} \text{d}t
\end{align}
The continuous extension is then obtained as:
\begin{align}
  Y \sim \text{Continuous Binomial}(n, p) & & &F_{Y}(y) = I_{1-p}(n-y, y+1)\\
  Z \sim \text{Continuous Negative Binomial}(r, p) & & &F_{Z}(z) = I_{1-p}(r, z+1)
\end{align}
These continuous extensions, result in an interpolation of both the
cumulative distribution and quantile functions of the discrete
counterparts. The advantage of this
interpolation scheme is that the behavior of the resulting continuous
random variables mimic that of their discrete counterparts. In the
discrete case it is well known that the Poisson distribution is the
limiting case of both the Binomial and the Negative Binomial and that
Binomial and Negative Binomial are entwined in a $1$-to-$1$ relation.
Theorem~\ref{theorem1} shows how these relationship between variables
are preserved in the continuous case, hence the two classes of
distribution have similar meaning. From a modeling perspective, in
fact, Theorem~\ref{theorem1} justifies the interpretation (1) of the
Continuous Poisson as an approximation for a Binomial-like
distribution in the case of rare events, (2) of the Continuous
Negative Binomial as an over-dispersed version of the Continuous
Poisson and (3) of the Continuous Negative Binomial as the waiting
time until the arrival of the $r$th success in a Binomial-like
experiment.
\begin{theorem}\label{theorem1}
    Let $X$ be a Continuous Poisson random variable with parameter
    $\lambda$, $Y $ be a Continuous Binomial with parameters $n$ and
    $p$ and $Z$ be a Continuous Negative Binomial with parameters $r$
    and $p$. Then the following relations hold:
    \begin{enumerate}
    \item for $n\rightarrow \infty$ and $p\rightarrow 0$ so that
        $np \rightarrow \lambda$
        \begin{equation}
            F_Y(x) = \frac{\text{B}(x+1, N-x, p)}{\text{B}(x+1, N-x)}
            \longrightarrow 
            \frac{\Gamma( x+1, \lambda)}{\Gamma(x+1)}  = F_X(x)
            \label{eq:bin_convergence}
        \end{equation}
    \item for $r\rightarrow \infty$ and $p\rightarrow 0$ so that
        $rp \rightarrow \lambda$ we have
        \begin{equation}
            F_Z(x) = \frac{\text{B}(x+1, r, p)}{\text{B}(x+1, r)}
            \longrightarrow 
            \frac{\Gamma( x+1, \lambda)}{\Gamma(x+1)}  = F_X(x)
            \label{eq:nbin_convergence}
        \end{equation}
    \item let $W$ be a Continuous Binomial random variable with
        parameters $s+r$ and $1-p$, then
        \begin{equation}
            F_Z(s)  = 1- F_W(r).
        \end{equation}
    \end{enumerate}
\end{theorem}

\section{An application - Disease Mapping}\label{sec:disease_mapping}

We reanalyse the hospitalisation data of \cite{Congdon2017}, described
in Table~\ref{tab:covariates}. The use of quantile regression instead
of mean regression is still relatively unexplored, with exceptions in
\cite{Congdon2017} and \cite{Chambers2014}. Since the focus of disease
mapping is on extreme behaviors of the population, however, quantiles
may be a more informative summary than means.
\begin{table}
 \centering
        \begin{tabular}{p{1.8cm}| p{10cm}}
          \textbf{Variable} & \textbf{Description} \\ \hline
          $Y_i$ & Counts of emergency hospitalizations for self-harm
                  collected in England over a period of \(5\) years
                  (from $2010$ to $2015$). The counts are
                  aggregated over \(6791\) are Middle Level Super Output Areas (MSOAs) \\
          $X_{{\tt Depr},i}$ & Deprivation, as measured by the 2015
                               Index of Multiple
                               Deprivation (IMD).\\
          $X_{{\tt SF},i}$ & Social fragmentation, measured by a
                             composite index derived from indicators
                             from the 2011 UK Census comprising
                             housing
                             condition and marital status.\\
          $X_{{\tt RS},i}$ & Rural status, again measured by a
                             composite indicator aimed at capturing
                             the accessibility to services and
                             facilities such as schools, doctors or
                             public offices.\\
                             \hline
        \end{tabular}
          \caption{Observed dataset composition}
    \label{tab:covariates}
\end{table}
Standard risk measures, such as the ratio between observed and
expected cases in each area, the Standardized Mortality (or Morbidity)
Ratio (SMR) \(SMR_i = { Y_i}/{ E_i}\), are not reliable here due to
the high variability of expected cases \( E_i\), hence is advisable to
introduce a random effect model that stabilizes the risk estimates by
borrowing information from the spatial structure of the data. Assuming
that, conditionally on covariates
$X_i = (X_{{\tt Depr},i}, X_{{\tt SF},i}, X_{{\tt RS},i})$ and random
effect \(b_i\), the observations are generated by a Poisson
distribution
\begin{equation}
    Y_i \mid  X_i, b_i \sim \text{Poisson}(\lambda_i) \quad (i = 1, \dots, n)
\end{equation}
we adopt the following model for the conditional quantile of level
\(\alpha\)
\begin{equation}\label{eq:relrisk}
    {Q}_{\alpha}({Y_i\mid X_i}, b_i) = 
    {E_i} \theta^{\alpha} _{i} = {  E_i} \exp\{\eta_i\} 
    \quad (i = 1, \dots, n).
\end{equation}
We opted for a \emph{quantile-level} approach for handling exposures
\(E_i\) in order to ease interpretation; as we discount each quantile
for the exposures, the parameter \(\theta_{i} ^{\alpha}\)
corresponding to the \(i^{\tt th}\) area can be considered the
relative risk of unit \(i\) at level \(\alpha\) of the population.
More details on the introduction of the exposures $E_i$ in the model
can be found in Appendix~\ref{app:exposures}. The linear predictor
\(\eta_i\) can be decomposed into
\begin{equation}
    \eta_i = \beta_0 + \beta_{\tt Depr} {  X}_{{\tt Depr},i} + 
    \beta_{\tt SF} {\tt X}_{{\tt SF},i} + 
    \beta_{\tt RS} {\tt X}_{{\tt RS},i} + b_i \quad (i = 1, \dots, n)
\end{equation}
where \(\beta_0\) represent the overall risk and \(b_i\) consists in
the sum of an unstructured random effect capturing overdispersion and
measurement errors and spatially structured random effect. In order to
avoid the confounding between the two components of the random effect
and to avoid scaling issues we adopt for \(b_i\) the modified version
of the Besag--York--Mollier (BYM) model introduced in
\cite{Simpson2017}:
\begin{equation}
    b_i = \frac{1}{\sqrt{\tau_b}}\left( \sqrt{1-\phi} v_i +
      \sqrt{\phi} u_i\right) \quad (i = 1, \dots, n).
\end{equation}
Both random effects are normally distributed, and in particular
\begin{align}
  v_i&\sim N(0, I) \quad (i = 1, \dots, n)\\
  u_i&\sim N(0, Q_u^{-1}) \quad (i = 1, \dots, n)
\end{align}
so that \(b_i\sim N(0 , Q_b ^{-1})\) with
\(Q_b^{-1} = \tau_b^{-1} ( 1-\phi)I + \phi Q_u^{-1}\), a weighted sum
of the identity matrix \({I}\) and the precision matrix representing
the spatial structure \(Q_u\), scaled in the sense of
\cite{Sorbye2014a}. We assign priors on the precision \(\tau_b\) and
the mixing parameter \(\phi\) using the penalized complexity (PC)
approach, as defined in \cite{Simpson2017} and detailed in
\cite{Riebler2016} in the special case of disease mapping.

Estimated coefficients shown in Table~\ref{tab:scotcoef} show that
Deprivation has a negative impact, which only slightly attenuates at
higher quantile level, meaning that, as we could expect, higher
deprivation corresponds to increases in self harm hospitalization.
Interestingly, being a rural area seems to have a positive effect
instead, with more rural areas being associated to lower rates of
hospitalization.

\begin{figure}
    \centering \includegraphics[width=1\textwidth]{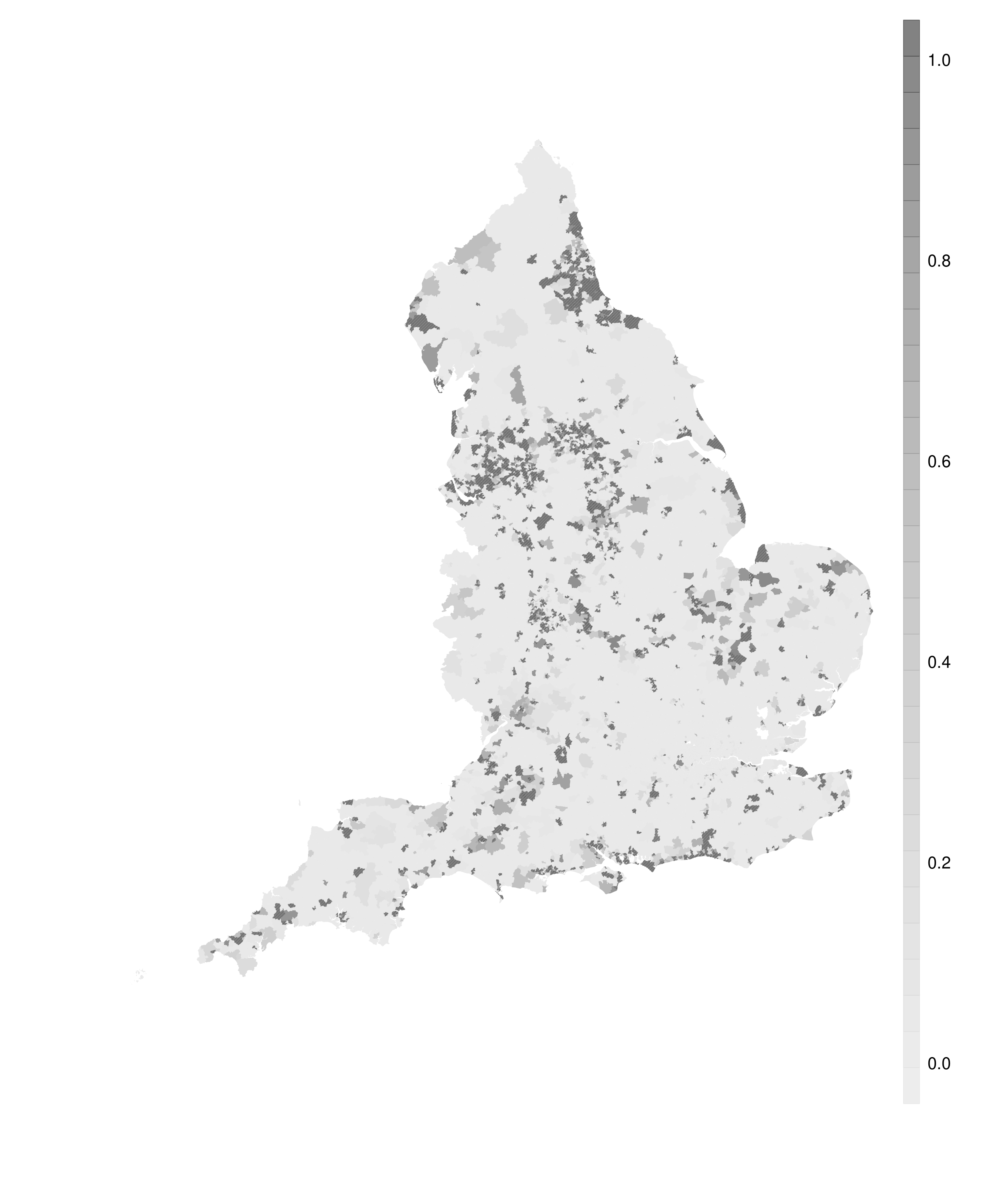}
    \caption{Exceeedance probability for Quantile Relative Risk. Dashed
        are areas of high risk.}
    \label{fig:ep}
\end{figure}

\begin{table}
\centering
        \begin{tabular}{l|cccc}
          & \textbf{Mean} & \textbf{1st Quartile} & \textbf{2nd Quartile} & \textbf{3rd Quartile} \\ \hline
          $\beta_0$          & -0.598 (0.015) & -0.707 (0.265) & -0.470 (0.275) & -0.512 (0.042) \\
          $\beta_{\tt Depr}$ & 1.981 (0.031) & 2.087 (0.220) & 1.960 (0.157) & 1.934 (0.059) \\
          $\beta_{\tt RS}$   & -0.814 (0.036) & -0.883 (0.127) & -0.878 (0.217) & -0.782 (0.217) \\
          $\beta_{\tt SF}$   & 0.429 (0.044) & 0.562 (0.155) & -0.098 (0.899) & 0.399 (0.105)\\
          $\tau_b$           & 6.409 (0.199) & 5.768 (0.180) & 6.274 (0.196) & 7.158 (0.177)\\
          $\phi$             & 0.838 (0.014) & 0.838 (0.014) & 0.838 (0.014) & 0.817 (0.011) \\ 
          \hline
        \end{tabular}
             \caption{Posterior mean estimates of model
        parameters (and corresponding standard deviations)}
    \label{tab:scotcoef}
\end{table}
The formulation of the model in terms of relative risk
$\theta_{i} ^{\alpha}$ as in~ \eqref{eq:relrisk}, is instrumental in
detecting areas of immediate concern. We identify the \(i\)th region
as ``high risk'' if the estimated probability of an increased relative
risk for the area is large enough, i.e. if \begin{equation}
    \text{pr}(\hat{\theta}_{i} ^{0.25}>1) > t \quad (i = 1, \dots, n)
\end{equation}
where $t$ is some user-defined threshold, typically $t = 0.9$ or
$t = 0.95$. While the use of exceeedance probabilities for relative
risk is common in the disease mapping literature, the benefit of our
proposal is to check for increases in the relative risk in the first
quartile, which is more worrisome than an increase in the average
relative risk. A risk map corresponding to $t = 0.95$ is shown in
Figure \ref{fig:ep}. Results are similar to those reported in
\cite{Congdon2017}, which defines the \(i\)th area region at ``high
risk'' if \([\hat{\theta}_{i, 0.05}, \hat{\theta}_{i,0.95}] >1\), thus
using only point estimates. With respect to this previous analysis,
the use of a model assumption on the response variable allows us to
make full use of the posterior distribution, resulting in a more
robust interpretation of high risk areas. Additionally, computing
exceeedance probabilities strengthens the role of quantile regression
in disease mapping.

\section*{Acknowledgment}

The authors would like to acknowledge Prof.\ Peter Congdon for
providing the MSOA hospitalization data.

\section{Appendix}
\subsection{Exposures}\label{app:exposures}

When count data depend on the size of the unit they have been observed
on, it is necessary to rescale them to allow for comparisons. There
are two ways of encoding a different observation window $E_i$ in the
model: by discounting the quantiles directly and
considering \[Q(Y_i|X_i) ^{\alpha}/E_i \quad (i = 1, \dots, n)\] or by
adjusting the global parameter and consider
\[\lambda_i/E_i \quad (i = 1, \dots, n).\]

While in Poisson mean regression these two approaches yield the same
results, in Poisson quantile regression they differ. In general
\begin{equation}
    \frac{\Gamma^{-1}( E_i \exp\{\eta_i ^{\alpha} \}
        ^{\alpha}+1, 1-\alpha)}{\Gamma( E_i \exp\{\eta_i ^{\alpha} \}
        ^{\alpha}+1)}\neq E_i \frac{\Gamma^{-1}(\exp\{\eta_i ^{\alpha}
        \} ^{\alpha}+1, 1-\alpha)}{\Gamma(\exp\{\eta_i ^{\alpha} \}
        ^{\alpha}+1)} \quad (i = 1, \dots, n).
\end{equation}
A case could be made for both modeling strategies, the former being a
``quantile-specific'' model while the latter being more of a global
model, and choosing between them depends on the application.

\subsection{Proof of Lemma 1}

By integration by parts we have
\begin{equation}
    \frac{\Gamma( x+1, \lambda)}{\Gamma(x+1)} - \frac{\Gamma( x, \lambda)}
    {\Gamma(x) }= \lambda^x e^{\lambda} /\Gamma(x+1)
\end{equation} which is enough to show that:
\begin{align}
  \text{pr}(\ceil{X}=x)  
  & =  \text{pr}(X\in (x-1, x])  = F_{X}(x)-F_{X}(x-1) \nonumber\\
  & = \frac{\Gamma( x+1, \lambda)}{\Gamma(x+1)} - \frac{\Gamma(x,
    \lambda)}{\Gamma(x)}
    = \lambda^x e^{\lambda} /\Gamma(x+1) \nonumber\\
  & = \text{pr}(X^* =x). \\
\end{align}

\subsection{Proof of Theorem 1}

\begin{enumerate}
\item immediately follows from \cite{Ilienko2013};
\item follows trivially from~\eqref{eq:bin_convergence};
\item follows from
    \begin{align}
      F_Z(s)  & = 1-I_p (s+1, r)\nonumber \\
              & = 1-I_p ((s+r) -(r-1), (r-1) +1) \nonumber\\
              & = 1 -  {P}(Y\leq r-1)\nonumber\\
              & =  {P}(Y\geq r).
    \end{align}
\end{enumerate}

\bibliographystyle{abbrvnat}
 \bibliography{paper-ref,mybib}
 \end{document}